\documentclass[useAMS,usenatbib]{mn2e}
\usepackage{graphicx,amssymb,ulem}
\usepackage[T1]{fontenc}
\usepackage{aecompl}
\usepackage{verbatim}
\usepackage{color}
\usepackage{eqnarray,amsmath}
\usepackage{multirow}
\usepackage{fixmath}
\usepackage{rotating}
\usepackage{pbox}
\usepackage{pdflscape}

\def\ap\deltajs{ApJS}

\title[Polarization in ultra-relativistic synchrotron sources]
  {Linear and Circular polarization in ultra-relativistic synchrotron sources -- implications to GRB afterglows}
\author[L. Nava et al.]
  {Lara Nava$^1$\thanks{lara.nava@mail.huji.ac.il},
  Ehud Nakar$^2$,
  Tsvi Piran$^1$
\\
 \\    
$^1$Racah Institute of Physics, The Hebrew University of Jerusalem, 91904, Israel\\
$^2$Raymond and Beverly Sackler School of Physics and Astronomy, Tel Aviv University, Tel Aviv 69978, Israel}
\pagerange{\pageref{firstpage}--\pageref{lastpage}} \pubyear{2015}
\begin{document}
\voffset -1truecm %astroph
\label{firstpage}

\maketitle

\begin{abstract}
Polarization measurements from relativistic outflows are a valuable tool to probe the geometry of the emission region and the microphysics of the particle distribution. Indeed, the polarization level depends on: (i) the local magnetic field orientation, (ii) the geometry of the emitting region with respect to the line of sight, and (iii) the electron pitch-angle distribution. Here we consider optically thin synchrotron emission and we extend the theory of circular polarization from a point source to an extended radially expanding relativistic jet. We present numerical estimates for both linear and circular polarization in such systems. We consider different configurations of the magnetic field, spherical and jetted outflows, isotropic and anisotropic pitch-angle distributions, and outline the difficulty in obtaining the reported high level of circular polarization observed in the afterglow of GRB 121024A. We conclude that the origin of the observed polarization cannot be intrinsic to an optically thin synchrotron process, even when the electron pitch-angle distribution is extremely anisotropic.
\end{abstract}


%=====================================================================================
%================================== INTRODUCTION =====================================
\section{Introduction}
Circular polarization at the level of $P^{circ}\sim0.6\%$ has been recently detected in the optical afterglow of GRB 121024A \citep{wiersema14}.
This is the first time that circular polarization is detected in GRB afterglow radiation.
The same burst also shows linear polarization at a level of $\sim4\%$, implying a ratio $P^{circ}/P^{lin}\simeq0.15$. 
Theoretical estimates of circular polarization in synchrotron radiation predict that both the absolute value of $P^{circ}$ and its value relative to the amount of linear polarization $P^{circ}/P^{lin}$ are $\lesssim1/\gamma_e$, where $\gamma_e$ is the random Lorentz factor of the radiating electrons \citep{legg68,sazonov69,sazonov72}. 
These calculations have been performed in the frame of the fluid, and for a point-like region where the magnetic field has a given orientation. 
Based on these order of magnitude estimates and assuming that synchrotron radiation is dominating the optical afterglow of GRB 121024A, the measured values of both $P^{circ}$ and $P^{circ}/P^{lin}$ are large as compared to expectations, and their theoretical interpretation is challenging.
For this burst, indeed, $\gamma_e$ is estimated to be of the order of $10^4$ at the time when observations are performed, implying that observations are well in excess of the predicted value.

A certain degree of linear polarization in GRB afterglow radiation is instead expected, and has been indeed measured in the optical afterglow of several GRBs (see \citealt{covino04} for a review).
In all cases, the reported values of $P^{lin}$ is at the level of a few percent, with the noticeable exception of GRB~120308A, where a higher level of polarization ($P^{lin}\simeq28\%$) has been measured in the early (four minutes after the prompt) optical afterglow \citep{mundell13}.
Besides being a confirmation of the synchrotron nature of the afterglow emission, the level of polarization has been used to infer the properties of the magnetic field configuration and the geometry of the emission. 
The level of polarization, indeed, depends on local conditions (as the orientation of the magnetic field and the properties of the electron population responsible for the emission) and on global conditions, as the geometry of the outflow.
For this reason, polarization measurements are in general considered a valuable tool to learn about the physics of the source. 

If the magnetic field is fully tangled and has the same strength in each directions then the radiation is unpolarised, no matter the geometry of the outflow. A symmetry break in the configuration of the magnetic field must be introduced in order to explain the observed polarization.
In the case of a spherical outflow, a completely tangled magnetic field in the plane of the shock (i.e. with no component in the direction perpendicular to the plane) would produce no net polarization, and some degree of coherence of magnetic field lines must be invoked in order to break the symmetry \citep{gruzinovwaxman99}. 
The situation is different in case of collimated outflows.
In this case in fact, even a completely random magnetic field in the plane of the shock can give rise to a net polarization, provided that the jet is seen off-axis  \citep{ghisellini99,sari99b}. 
This model predicts a change by $90^\circ$ in the polarization position angle around the jet break time and a typical evolution of the degree of linear polarization.
Different predictions on the evolution of $P^{lin}$ and its position angle are derived if an ordered component and/or a component perpendicular to the plane of the shock is considered, \citep{granotkonigl03,granot03,nakar03} or if the magnetic field is significantly different in the parallel and perpendicular directions and sideways spreading of the jet is considered \citep{sari99b}.
The temporal evolution of the linear polarization has also been proposed as a tool to reveal the structure of the jet, based on the fact that polarization curves for structured jets are very different from those derived for uniform jets \citep{rossi04}. 

While in general observations of linear polarization can be explained within the current theoretical models, the detection of circular polarization was an unexpected result.
In the original paper where the detection is presented \citep{wiersema14}, the authors show that it is very unlikely that its origin could be ascribed to plasma propagation effects or dust scattering, and argue in favour of an intrinsic origin. In particular, they invoke an anisotropic distribution of the electron pitch angles as a possible explanation.
However, estimates are performed in the local frame of the fluid and do not take into account that the contribution of different emitting regions to the total polarization must be averaged, in order to derive the total (integrated over the emitting region) polarization in the frame of the observer.
The relativistic motion of the outflow (that requires application of Lorentz transformations) and the integration over the (unresolved) image both play an important role in the determination of the final polarization detected by the observer and should be properly treated.

In this work we derive the circular polarization of synchrotron radiation from a relativistic outflow. The amount of linear polarization for different magnetic field configurations and outflow geometries has been already derived in previous works \citep{sari99b,ghisellini99,gruzinovwaxman99,gruzinov99,granotkonigl03,granot03,nakar03,rossi04}. However, since we are also interested in estimating the ratio $P^{circ}/P^{lin}$, we also report the equations to derive the linear polarization.

The paper is organised as follows. In \S\ref{sect:origin} we provide an order of magnitude estimate and a schematic description of the origin of polarization in synchrotron radiation. 
Section \S\ref{sect:lsp} introduces the Stokes parameters in the comoving frame (i.e. the frame at rest with the source). In \S\ref{sect:configuration} the geometrical set-up we are considering is presented. Lorentz transformation to move from the comoving to the observer frame are presented in \S\ref{sect:lt}. The equations to derive the total polarization from an unresolved source are derived in  \S\ref{sect:wf}. In \S\ref{sect:results} we estimate the circular and linear polarization for different configurations. We discuss our results in comparison with polarization measurements of the optical afterglow of GRB121024A in section \S\ref{sect:discussion}. Finally, in \S\ref{sect:conclusions} we summarise our conclusions.

%========================================================================================
%======================================= THE ORIGIN =======================================
\section{The origin of polarization in synchrotron radiation}
\label{sect:origin}
In this section we discuss what is the origin of polarization in synchrotron radiation and why the level of circular polarization is expected to be small as compared to the level of linear polarization. This discussion is also aimed at understanding which are the physical parameters the determine the expected level of polarization in the comoving frame of the fluid.
%------------------------------------------------------
\begin{figure}
\vskip -4.5 truecm
\hskip -1.5truecm
\includegraphics[scale=0.7]{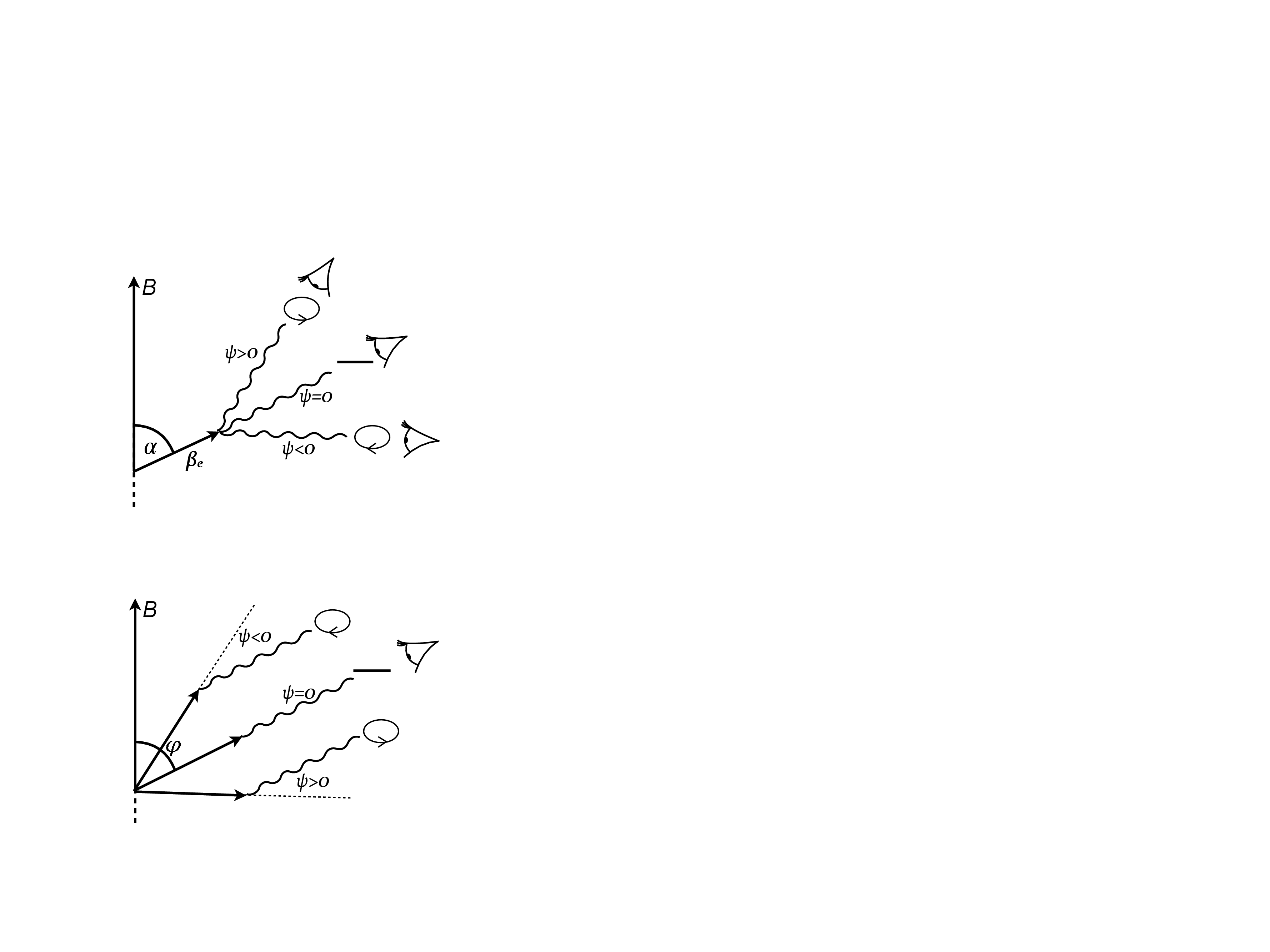}
\vskip -1.8 truecm
\caption{Upper panel: polarization of the radiation emitted by an electron with pitch angle $\alpha$ as detected by different observes. The angle $\psi$ represents the angle between the electron velocity $\mathbold{\beta_e}$ and the direction of the photon. Lower panel: polarization from a population of electrons with different pitch angles, as detected by one single observer. $\varphi$ is the angle between the magnetic field and the direction of the observer.}
\label{fig:sketch2}
\end{figure}
%------------------------------------------------------

We first consider emission from one single electron with a pitch angle $\alpha$, gyrating in a uniform magnetic field ${\mathbold B}$ (Figure~\ref{fig:sketch2}, upper panel). 
Let $\psi$ be the angle between the electron velocity ${\mathbold \beta_e}$ and the direction of the photon ${\mathbold n}$, and let consider three different observers detecting radiation from the same electron.
In general, the observed radiation is elliptically polarised, with the axes of the ellipse parallel and perpendicular to the projection of ${\mathbold B}$ on to the plane transverse to the direction of the photon ${\mathbold n}$ \citep{westfold59,legg68}. 
The major axis of the polarization ellipse is perpendicular to the projection of ${\mathbold B}$ for $|\psi|$ close to zero, but as $|\psi|$ increases, the form of the ellipse varies to a circle and then to an ellipse with major axis parallel to the projection of ${\mathbold B}$. 
The direction is right-handed ($RH$) or left-handed ($LH$) according with $\psi\gtrless0$. The polarization is linear if $\psi=0$.
We note a negligible amount of radiation is radiated at angles $|\psi|>1/\gamma_e=\sqrt{1-\beta_e^2}$. 

Let us now consider a population of electrons with different pitch angles, and a single observer, located at an angle $\varphi$ from the magnetic field (Figure~\ref{fig:sketch2}, lower panel).  As follows from the previous discussion, photons corresponding to $\psi=0$ (i.e., emitted in the direction of the observer by electrons with $\alpha=\varphi$) are linearly polarized. Photons for which $\psi\neq0$ instead will be elliptically polarised. The elliptical polarization is the same for photons at $\psi$ and $-\psi$, but with opposite direction of the rotation. 

The resulting polarization detected by the observer is obtained after integration over the contributions from all photons that reach the observer. 
First we note that only electrons with pitch angle $\varphi-1/\gamma_e<\alpha<\varphi+1/\gamma_e$ give a non negligible contribution to the emission detected by the observer. 
These pitch angles correspond to photons with very small $\psi$. As noticed before, small $\psi$ correspond to polarization ellipses with major axes perpendicular to the projection of $\mathbold{B}$ on the plane perpendicular to the observer. 
This implies that when the contribution from all these ellipses with all major axes perpendicular to the projection of $\mathbold{B}$ is summed up the final level of integrated lineal polarization is still large (around 70\%). 
On the contrary, the integrated circular polarization, is largely reduced by integration.
Indeed, the contribution from photons with $\psi>0$ is partially cancelled by the contribution from photons with $\psi<0$, since different signs of $\psi$ correspond to different directions of the rotation.
This cancellation suppresses the final amount of circular polarization, which is expected to be small as compared to the linear polarization. 
The level of cancellation depends on how many photons have positive and negative $\psi$: if the number of photons with angle $\psi$ is equal to the number of photons with angle $-\psi$ no net circular polarization is detected.
The number of photons at different angles depends on the number of electrons at different pith angles.
This is the reason why, besides the angle $\varphi$, the properties of the polarization obtained after averaging over the contributions from different electrons depend also on the pitch angle distribution and on the shape of the electron energy spectrum. 

In the following we consider an electron distribution that has a power-law dependence on the energy, the number of electrons with Lorentz factor $\gamma_e$ per unit solid angle around the pitch angle $\alpha$ is:
%------------------------------------------------------
\begin{equation}
N(\alpha,\gamma_e)\propto Y(\alpha)\gamma_e^{-\delta}.
\end{equation}
%------------------------------------------------------
With the index $\delta$ we refer to the electron distribution after radiative cooling has modified the initial injected distribution.
Hence this distribution can be different than the injected electron spectrum $N_{inj}(\gamma_e)\propto \gamma_e^{-p}$. For an uncooled distribution $\delta=p$, while for cooled electrons $\delta=p+1$ or $\delta=2$, depending on the value of $\gamma_e$ as compared to the minimum and the cooling Lorentz factors.

The function $Y(\alpha)$ is defined as:
%------------------------------------------------------
\begin{equation}
Y(\alpha)=\frac{dN(E,\alpha)}{d\Omega}\frac{4\pi}{N_e(E)}.
\end{equation}
%------------------------------------------------------
where $N_e(E)$ is the total number density of electrons with energy $E$. Following this definition, $Y=1$ for an isotropic pitch angle distribution.

The influence of the pitch angle distribution is clear (see Figure~\ref{fig:sketch2}): if the number of electrons with pitch angle $\alpha=\varphi+\psi$ is different from the number of electrons with pitch angle $\alpha=\varphi-\psi$, the contributions to the total polarization from electrons at $\psi$ and $-\psi$ do not cancel out. 
For example, for the case of isotropic pitch angle distribution, if $\varphi<90^\circ$ the number of electrons with $\psi>0$ is larger than the number of electrons with $\psi<0$ and the total polarization is negative (as in the case in Figure~\ref{fig:sketch2}). Viceversa, if $\varphi>90^\circ$ the total polarization is positive.

The role played by the power-law index $\delta$ of the Lorentz factor distribution can be understood by first noticing that for a given fixed frequency $\gamma_e^2\propto1/\sin{\alpha}$. Electrons with different pitch angles (as the ones represented in Figure~\ref{fig:sketch2}) must then have different Lorentz factors in order to radiate photons at the relevant frequency $\nu$. For this reason the overall polarization is also determined by the number of electrons at different $\gamma_e$, i.e. from $\delta$.

%=========================================================================================================
%==================================== LOCAL STOKES PARAMETERS ===========================================

\section{Local Stokes parameters}
\label{sect:lsp}
The Stokes parameters $I, Q, U, V$ describe the polarization state of the electromagnetic radiation. They represent ac- tual intensities, and more specifically, linear combinations of intensities measured in orthogonal polarizations directions. In terms of the parameters of the polarization ellipse (that has been qualitatively described in Section 2), the Stokes parameters for a monochromatic wave are:
\begin{align}
I & = (a^2+b^2), \\
Q & =  (a^2-b^2)\cos{2\lambda}, \nonumber \\
U & =  (a^2-b^2)\sin{2\lambda}, \nonumber \\
V & =  \pm 2ab \nonumber
\end{align}
where $a$ and $b$ refer to the semi-major and semi-minor axes of the ellipse, and $\lambda$ is the tilt angle of the ellipse with respect to some reference direction.
The parameter $I$ describes the total intensity of the beam, $Q$ describes the excess of linearly horizontally polarized light over linearly vertically polarized light, $U$ describes the excess of linear $+45^\circ$ polarized light over linear $-45^\circ$ polarized light, and the fourth parameter $V$ describes the excess of right-circularly polarized light over left-circularly polarized light.

As explained before, the major and minor axis of the polarization ellipse (and then, the Stokes parameters) depend on the angle $\psi$ between the photon direction and the electron velocity. Then integration over all the radiation must be performed in order to find the Stokes parameters coming from a population of electrons, located at some positions of the source. 
In the next sections we give the equations for the Stokes parameters averaged over the electron population, in the comoving frame of the source.
We add the subscript '0' in order to distinguish these local Stokes parameters (i.e. estimated for a point-like region) from those averaged over the spatially extended emission region.
Since the linear and circular polarization are a combination of the {\it normalised} Stokes parameters  (i.e. relative to the total intensity of the beam, described by the first parameter $I_0$), in the following we give te equations for $U_0/I_0$, $Q_0/I_0$, and $V_0/I_0$.

%==================================  LINEAR POLARIZATION  =====================================
\subsection{Linear polarization: the Stokes parameters $Q_0/I_0$ and $U_0/I_0$}
The linear polarization $P^{lin}_{max}$ from an isotropic electron distribution, and for a given direction of the magnetic field is \citep{rybicki86}:
%----------------------------------------------------
\begin{equation}
P^{lin}_{max}=(\delta+1)/(\delta+7/3).
\label{eq:plinmax}
\end{equation}
%----------------------------------------------------
This quantity can also be expressed in terms of the spectral index $s$ of the radiation spectrum (defined by $dF_\nu/dE\propto E^{-s}$). 
Recalling that the spectral index of radiation emitted by a population of electrons with a power-law energy spectrum ($dN_e/dE\propto E^{-\delta}$) is given by$s=(\delta-1)/2$, we obtain $P^{lin}_{max}=(s+1)/(s+5/3)$.

For configurations where locally the magnetic field is ordered (like uniform in the plane of the shock, or perpendicular to the plane of the shock), 
the local linear polarization is:
%----------------------------------------------------
\begin{equation}
P^{lin}_0=P^{lin}_{max}~~~~~~(\rm orderd),
\label{eq:plin0_ord}
\end{equation}
%----------------------------------------------------
and does not depend on the position. For typical values of $\delta$, the linear polarization estimated from a point-like region is very high (as predicted in the quantitative description in Section~\ref{sect:origin}), and is around $\sim60$-$70\%$. 

For configurations where the magnetic field is locally tangled, the polarization is obtained after averaging over all possible directions of the magnetic field. In the case of a perfectly random magnetic field in the plane of the shock \citep{sari99b,granotkonigl03}:
%----------------------------------------------------
\begin{equation}
P^{lin}_0=P^{lin}_{max}\frac{\sin\theta^\prime}{1+\cos^2\theta^\prime}~~~~~~(\rm random~in~the~plane).
\label{eq:plin0_random}
\end{equation}
%----------------------------------------------------
The angle $\theta^\prime$ is the angle between the direction of the photon (in the comoving frame) and the normal to the plane where the magnetic field lies. 
In this case the local polarization depends on the position of the observer as compared to the plane of the shock.

The linear polarization is described by $Q_0/I_0$ and $U_0/I_0$.
%----------------------------------------------------
\begin{eqnarray}
Q_0/I_0=P^{lin}_0\cos(2\theta_p)\label{eq:qu0}\\  
U_0/I_0=P^{lin}_0\sin(2\theta_p)\nonumber 
\end{eqnarray}
%----------------------------------------------------
The angle $\theta_p$ is the position angle and gives the orientation of the linear polarization in some coordinate system.
In configurations with ordered magnetic field, even if the local polarization is independent on the location of the source (see equation~\ref{eq:plin0_ord}), the polarization angle is not, and when the contribution from different regions is summed partial cancellation can arise, depending on the geometry of the system.

%================================== CIRCULAR POLARIZATION =====================================
\subsection{Circular polarization: the Stokes parameter $V_0/I_0$}
\label{sect:circ_pol}
The Stokes parameter $V_0/I_0$ describes the level of circular polarization.
Its derivation for a power-law distribution of electrons is given by \citep{sazonov69,sazonov72}:

%--------------------------------------------------
\begin{eqnarray}
P^{circ}_{0}\equiv\frac{V_0}{I_0}=-\frac{4\sqrt{2}(\delta+1)(\delta+2)}{3\delta(\delta+7/3)}\frac{\Gamma\left(\frac{3\delta+8}{12}\right)\Gamma\left(\frac{3\delta+4}{12}\right)} {\Gamma\left(\frac{3\delta+7}{12}\right)\Gamma\left(\frac{3\delta-1}{12}\right)}\times \nonumber\\
\times \left(\cot{\varphi}+\frac{1}{\delta+2}\frac{1}{Y(\varphi)}\frac{dY(\varphi)}{d\varphi}\right)\frac{1}{\gamma_e}.
\label{eq:p0circ}
\end{eqnarray}
%----------------------------------------------------
This equation is valid provided that $|\frac{1}{Y(\varphi)}\frac{dY(\varphi)}{d\varphi}|< \gamma_e$. The circular polarization can also be expressed in terms of the linear polarization $P^{lin}_0=P^{lin}_{max}$:
%----------------------------------------------------
\begin{equation}
P^{circ}_{0}\simeq -\frac{P^{lin}_{max}}{\gamma_e} \times \left(\cot{\varphi}+\frac{1}{\delta+2}\frac{1}{Y(\varphi)}\frac{dY(\varphi)}{d\varphi}\right),
\label{eq:pcirc/plin}
\end{equation}
%----------------------------------------------------
where we have neglected factors order unity. 

In the standard case of isotropic pitch angle distribution (where the second term of the sum in brackets vanishes) the ratio between circular and linear polarization is of the order of $1/\gamma_e\ll1$, and since $P^{lin}_{max}\sim 0.7$ the value of $P^{circ}_0$ is also of the order  $1/\gamma_e\ll1$. Locally the circular polarization is strongly suppressed  and is of the order $1/\gamma_e\ll1$. As anticipated in Section~\ref{sect:origin}, the reason for this suppression is the cancellation between electrons contributing with positive and negative polarization.
A high level of circular polarization can be obtained only from an electron population with a high level of pitch angle anisotropy.
However, this is not the end of the story.
Equation \ref{eq:pcirc/plin} in fact cannot be compared directly to the observations. Comparison with observations requires that we first 
transform from the comoving to the observer frame and average polarization coming from different regions, as it will become clear in the next sections.

%==================================== THE GEOMETRICAL SETUP ===========================================
\section{The geometrical set-up}
\label{sect:configuration}
%------------------------------------------------------
\begin{figure}
\vskip -1.5 truecm
\hskip -2.8truecm
\includegraphics[scale=0.37]{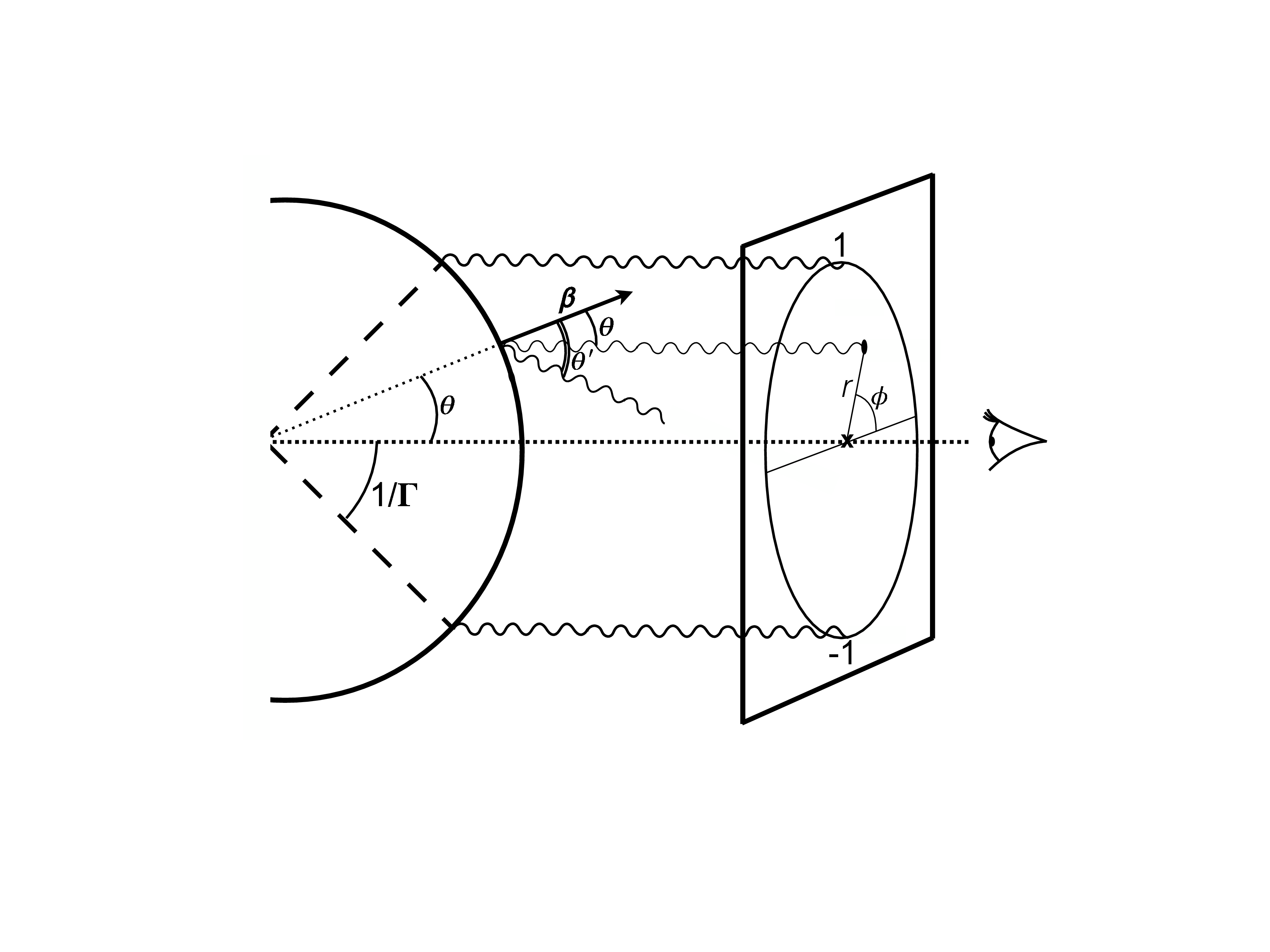}
\vskip -2 truecm
\caption{Representation of the geometrical set-up. In the frame at rest with the fluid, the photon is emitted at an angle $\theta^{\prime}$ from the direction of the fluid velocity $\beta$, but in the laboratory frame it is beamed in the direction of the observer. The emitted radiation is then mapped onto the plane of the sky, where we introduce a system of polar coordinates $(r,\phi)$. The coordinate $r$ is equal to unity for a region located at $\theta=1/\Gamma$. The contribution from radiation at $r>1$ is negligible.}
\label{fig:sketch}
\end{figure}
%------------------------------------------------------
In the previous section we have discussed polarization in the comoving frame, coming from electrons located in a specific point-like region (i.e. a region where the magnetic field has a given orientation). However, we are interested in deriving polarization from a blast wave expanding at relativistic velocity $\Gamma\gg1$. This has two implications: i) we have a spatially extended emitting region that, as we will discuss, is characterised in each point by different local conditions, and ii) the system discussed in the previous section is now moving at relativistic speed with respect to the laboratory frame.
In what follows, we describe the geometrical set-up that we are considering. A sketch is depicted in Figure~\ref{fig:sketch}.
In this sketch the blast wave is represented as a sphere (or, equivalently a jet with jet opening angle $\theta_{jet}\gg1/\Gamma$ and edges that are not visible to the observer). 
However, the procedure that we will present is also valid for jets with $\theta_{jet}\lesssim1/\Gamma$ and/or jets seen off axis.

In the comoving frame, the direction of the photon is identified by the angle $\theta^\prime$, measured from the direction of the bulk motion $\mathbold{\beta}=\sqrt{\Gamma^2-1}/\Gamma$. 
Because of beaming, in the observer frame this photon is boosted in the direction of the observer, and makes an angle $\theta$ with the fluid velocity $\mathbold \beta$ (see Figure~\ref{fig:sketch}). The relation between the two angles in the two different frames is given by a Lorentz transformation, and is described in Section~\ref{sect:lt}.
Besides the direction of the photon in the comoving frame, we also need the direction of the magnetic field in order to compute the polarization. This depends on the configuration of the field.
However, it can be immediately realised that both the field component in the plane of the shock (i.e., tangential to the sphere surface) and the field component perpendicular to the plane of the shock (i.e., parallel to the radial velocity $\mathbold\beta$) have different orientations at each point of the emitting surface.
From this discussion it is clear that the observer will detect different polarization from different regions.

The procedure to compute the resulting polarization detected by the observer (i.e., the quantity that must be compared to the observational measurements), is the following.
First, one needs to estimate the local (i.e. in a small region) polarization (that we have described in terms of local Stokes parameters) in the comoving frame of the source (Section~\ref{sect:lsp}). Each different region will be characterised by a different polarization status. Then one needs to apply Lorentz transformation to move from the comoving frame to the observer frame (Section~\ref{sect:lt}). Since the region from which the radiation originates is spatially extended, these different regions and their local polarization status are mapped into a plane perpendicular to the line of sight, i.e. the plane of the sky, and result in a polarization map (see Figure~\ref{fig:sketch}). The emission from GRB afterglows, however, is spatially unresolved, so that the detected polarization comes from the integration over this polarization map. To estimate this integrated polarization we first need to average each Stokes parameter. The average over the plane of the sky is, of course, weighted by the intensity of the emission. Finally, from the averaged Stokes parameters, the total integrated polarization is obtained (Section~\ref{sect:wf}).

In order to describe the local Stokes parameters in each point of the plane of the sky, we introduce a system of polar coordinates ($r,\phi$) on this plane.
The center of this coordinate system is located at the intersection between the plane of the sky and the direction connecting the observer and the central engine.
We define  $r\equiv\sin\theta\Gamma\simeq\theta\Gamma$ (for $r\ll\Gamma$), so that $r=1$ corresponds to radiation coming from a region located at $\sin{\theta}=1/\Gamma$. 
This is the visible region for the observer: electrons travelling at angles $\theta>1/\Gamma$ do not significantly contribute to the radiation received by this observer.
We will also use the coordinate $y$ in place of $r$, defined as $y\equiv(\theta\Gamma)^2=r^2$. We call polarization map the mapping of the polarization (in the lab frame) coming from different locations of the surface. We will give examples of polarization maps in section \ref{sect:results}.

%====================== LORENTZ TRANSFORMATIONS ==========================
\section{The Lorentz transformations}
\label{sect:lt}
In order to explicitly estimate the local Stokes parameters, we first need to express the angles $\theta_p$ and $\varphi$ and the Lorentz factor $\gamma_e$ as a function of quantities in the observer frame.
We apply these transformations to each point of the emitting region and give the relevant parameters $\theta_p$, $\varphi$, and $\gamma_e$ as a function of the polar coordinate system $(r,\phi)$ introduced in the plane of the sky.

We are interested in those photons that, after Lorentz boosting, travel towards the observer. These photons are those emitted at an angle $\theta$ from the direction of the fluid velocity $\mathbold{\beta}$. In the local frame (primed quantities), this angle is:
%----------------------------------------------------
\begin{equation}
\sin{\theta^\prime}=\mathcal{D} \sin{\theta}=\frac{\sin{\theta}}{\Gamma(1-\beta\cos{\theta})},
\end{equation}
where $\mathcal{D}$ is the Doppler factor:
%----------------------------------------------------
\begin{equation}
\mathcal{D}=\frac{1}{\Gamma(1-\beta\cos{\theta})}=\frac{2\Gamma}{1+y}.
\end{equation}
%----------------------------------------------------
This equation (where we have used the approximation $\cos\theta\simeq1-\theta^2/2$) shows that the Doppler factor depends on the position on the map.
This is due to the fact that we are considering a spherical emitting surface (or a part of a sphere, in case of jetted outflows), where the fluid moves in radial directions. 
This implies that different parts of the fluid move at different angles $\theta$ with respect to the line of sight. 
Since $\sin\theta\simeq\theta$ we can write the angle $\theta^\prime$ (that identifies the direction of the photon in the local frame) as a function of the coordinates in the laboratory frame:
%----------------------------------------------------
\begin{equation}
\sin\theta^\prime=\frac{2\sqrt{y}}{1+y}.
\label{eq:sinthetap}
\end{equation}
%----------------------------------------------------

Now to estimate $\theta_p$ we proceed as follows. First we need to identify in each point the direction of the magnetic field. This will depend on the configuration considered. Once the configuration has been chosen, the polarization vector in the local frame can be derived from the vectorial product between the magnetic field vector and the direction of the photon. Now that the polarization vector has been identified, it is necessary to move back to the laboratory frame and describe the orientation of the polarization vector in our coordinate system. 
We give the polarization angle for the three different magnetic field configurations that we explore in the following: i) uniform in the plane of the shock, ii) random in the plane of the shock, and iii) radial, i.e., perpendicular to the plane of the shock. For case i) we consider a magnetic field parallel to the axis $\phi=0$:
%-------------------------------------------------
\begin{eqnarray}
\theta_p=\phi+\arctan\left(\frac{1-y}{1+y}\cot{\phi}\right) ~~~~~~\rm (uniform),\\
\theta_p=\phi ~~~~~~~~\rm (random),\\
\theta_p=\phi+90^\circ ~~~~~\rm (radial).
\end{eqnarray}
%--------------------------------------------------
These equations for $\theta_p$, together with equations \ref{eq:qu0} and \ref{eq:plinmax} allows us to compute the normalised Stokes parameters $Q_0/I_0$ and $U_0/I_0$ in the laboratory frame at each position on the plane of the sky. Note that $P^{lin}_0$ is given by equation~\ref{eq:plin0_ord} for locally ordered magnetic fields (namely, uniform in the plane of the shock or radial), while is given by equation~\ref{eq:plin0_random} for a random magnetic field. In this last case, using equation~\ref{eq:sinthetap} we can write the local linear polarization as a function of the coordinate $y$: $P^{lin}_0=P^{lin}_{max}\times y/(1+y^2)$.

To derive $V_0/I_0$ as a function of the position on the plane of the sky, we need to calculate $\varphi(y,\phi)$ and $\gamma_e(y,\phi)$.
To derive $\gamma_e$ we first note that the observed frequency is $\nu^{obs}=\nu/(1+z)=\nu^\prime\mathcal{D}/(1+z)$, where the Doppler factor $\mathcal{D}$ is to be applied to transform from the comoving frame frequency $\nu^\prime$ to the progenitor frame frequency $\nu$. 

The comoving frequency $\nu^\prime$ is given be synchrotron frequency formula:
%-------------------------------------------------
\begin{equation}
\nu^{\prime}=\frac{3\nu_H\sin{\alpha}}{2}\gamma^2_e,
\label{eq:nuprime}
\end{equation}
%-------------------------------------------------
where $\nu_H=eB/(2\pi m_ec)$.
The pitch angle $\alpha$ can be taken equal to $\varphi$, since in deriving equation~\ref{eq:p0circ} the average on the electrons with slightly different pitch angles (in the cone $1/\gamma_e$) has been already performed. 
By inverting equation \ref{eq:nuprime} and writing everything in terms of the observed frequency we obtain the Lorentz factor of the electrons that radiate at the relevant frequency:
%-------------------------------------------------
\begin{equation}
\gamma_e^2=\frac{\nu^{obs}(1+z)(1+y)}{3\nu_H\sin{\varphi}\Gamma}.
\label{eq:ge}
\end{equation}
%-------------------------------------------------
Equation~\ref{eq:ge} shows that the contribution to the emission at a given observed frequency $\nu^{obs}$ (at which polarization measurements are performed) comes from electrons with different $\gamma_e$.
This is due to the fact that in order to move from the observer to the comoving frame we need to account for the Doppler factor $\mathcal{D}(y)$ and that the relevant pitch angle of those electrons which are radiating in the direction of the observer also varies.

Finally we need to estimate $\varphi$. This is the angle between the photon direction and the field lines, and it will depend on the configuration of the magnetic field. For a given direction of the magnetic field, the angle $\varphi$ is obtained by estimating the scalar product (in the local frame) between the photon direction and the magnetic field.
We give its value for a uniform magnetic field in the plane of the shock and for a radial magnetic field (in a random magnetic field the local circular polarization is always zero):
\begin{eqnarray}
\sin^2{\varphi} = \left(\frac{1-y}{1+y}\right)^2\cos^2\phi+\sin^2{\phi} ~~~~~~~~\rm (uniform)\label{eq:varphi_unif},  \\
\sin\varphi=\frac{2\sqrt{y}}{1+y} ~~~~~~~~~~~~\rm (radial)\label{eq:varphi_rad}.
\end{eqnarray}

%====================================  TOTAL POLARIZATION  =======================================
\section{The total polarization}
\label{sect:wf}
To derive the total polarization we integrate each Stokes parameter separately over its map on the plane of the sky, after the contribution coming from different regions has been weighted by the intensity. Namely, the value of the Stokes parameter in each point of the map is weighted by the ratio between the contribution of that region to the total flux and the total flux.

We define $\frac{dF_\nu(y,\phi)}{dS}$ as the specific flux per unit surface $dS=d\phi rdr=d\phi dy/2$ coming from a region located at $(y,\phi)$.
Its value is proportional to: $\frac{dF_\nu(y,\phi)}{dS}\propto I_\nu d\Omega\propto \mathcal{D}^3 I^\prime_{\nu^\prime}$. Recalling that the specific intensity $I^\prime_{\nu^\prime}$ in the comoving frame of the fluid is proportional to the number of electrons pointing in the direction of the observer we obtain:
%------------------------------------------------
\begin{align}
\frac{dF_\nu(y,\phi)}{dS}&\propto Y(\varphi)\mathcal{D}^3 (\nu^\prime)^{-s}(B\sin{\varphi})^{s+1}\propto Y(\varphi)\mathcal{D}^{3+s}(B\sin{\varphi})^{s+1}\nonumber\\
&\propto Y(\varphi)(1+y)^{-(3+s)}(\sin{\varphi})^{s+1}.
\label{eq:wf}
\end{align}
%------------------------------------------------

The total flux is the integral of equation \eqref{eq:wf} on the region of the map that corresponds to the emitting surface $dS$:
%-------------------------------------------------
\begin{equation}
F_\nu=\int \frac{dF_\nu(y,\phi)}{dS} dS\propto \int Y(\varphi)(1+y)^{-(3+s)}(\sin{\varphi})^{s+1} d\phi dy/2.
\label{eq:totflux}
\end{equation}
%-------------------------------------------------
The proportionality constants in equations \ref{eq:wf} and \ref{eq:totflux} are the same, and we don't need to explicitly write them, since we only need the ratio between the partial flux at a given position $\frac{dF_\nu(y,\phi)}{dS}$ and the total integrated flux $F_\nu$.
Integration must be performed at least up to $y\gtrsim1$, since this is the region where most of the emissivity come from. Integration up to larger distances can introduce differences at the level of $10\%$ \citep{granot03}.

If the outflow instead of being spherical is collimated into a jet, in the plane of the sky the jet can be described in terms of position of its axis $(r_{jet},\phi_{jet})$ and opening angle $\theta_{jet}$.  Note that $r_{jet}=\theta_{view}\Gamma$, where $\theta_{view}$ is the viewing angle, i.e. the angle between the line of sight and the jet axis.
For on-axis jets, $r_{jet}=0$ and $\phi_{jet}=0$, and the edges of the jet are described by $r^2=\theta_{jet}\Gamma$. Outside this region the flux vanishes. 
More generally, for off-axis jets $F_\nu(y,\phi)=0$ in the region of the plane that satisfies the condition: $(r\cos\phi-r_{jet}\cos\phi_{jet})^2+(r\sin\phi-r_{jet}\sin\phi_{jet})^2>\theta_{jet}\Gamma$.

Taking as an example $U_0/I_0$ (but the same equations can be applied to the other normalised Stokes parameters) the flux-weighted parameter is given by:
%--------------------------------------------------------------------
\begin{align}
\left(\frac{U_0}{I_0}\right)_{fw}&\equiv\frac{U_0}{I_0}\times\frac{\frac{dF_\nu(y,\phi)}{dS}}{\int \frac{dF_\nu(y,\phi)}{dS}dS}= \nonumber \\
&=\frac{U_0}{I_0}\times \frac{Y(\varphi)(1+y)^{-(3+s)}(\sin{\varphi})^{s+1}}{\int Y(\varphi)(1+y)^{-(3+s)}(\sin{\varphi})^{s+1} dS} 
\label{eq:fwsp}
\end{align}
%--------------------------------------------------------------------
The subscript $fw$ stands for 'flux-weighted'. We recall that the normalised Stokes parameters are also a function of the position $(y, \phi)$. 
The integration of the quantity in equation~ \ref{eq:fwsp} over all the surface will give the Stokes parameters for an unresolved source:
%---------------------------------------------------------------
\begin{equation}
\frac{U}{I}=\int  \left(\frac{U_0}{I_0}\right)_{fw} dS = \frac{\int \frac{U_0}{I_0}\times \frac{dF_\nu(y,\phi)}{dS}d\phi dy}{\int \frac{dF_\nu(y,\phi)}{dS}d\phi dy}
\label{eq:swf}
\end{equation}
%---------------------------------------------------------------

The total linear polarization is given by 
\begin{equation}
P^{lin}=\sqrt{ \left(\frac{Q}{I}\right)^2+\left(\frac{U}{I}\right)^2}
\end{equation}

The total circular polarization is given by:
\begin{equation}
P^{circ}=\frac{V}{I}=\frac{\int{\frac{V_0}{I_0}\frac{dF_\nu(y,\phi)}{dS} dS}}{\int{\frac{dF_\nu(y,\phi)}{dS}dS}}
\end{equation}

%============================================================================================
%====================================  RESULTS: SOME EXAMPLE ======================================
\section{Results: some examples}
\label{sect:results}
Using the above formalism we discuss now polarization estimates for several configurations. 
For the geometry of the outflow we consider three different cases: 
\begin{itemize}
\item a sphere (or equivalently a jet with jet opening angle $\theta_{jet}\gg1/\Gamma$),
\item a narrow jet with $\theta_{jet}=1/\Gamma$,
\item a narrow jet with $\theta_{jet}=1/3\Gamma$.
\end{itemize}
For the jetted geometries, the results depend on the viewing angle $\theta_{view}$ between the jet axis and the line of sight. 

For the pitch angle distribution we consider:
\begin{itemize}
\item the standard isotropic pitch angle distribution (corresponding to $Y(\alpha)=1$),
\item anisotropic distributions.
\end{itemize}

We consider configurations of the magnetic field where the  symmetry is broken, since completely tangled magnetic fields do not give rise to net polarization. We consider:
\begin{itemize}
\item a random filed in the plane of the shock (i.e., with no component in the radial direction),
\item a radial magnetic field (i.e., with no component in the plane of the shock),
\item a uniform magnetic field in the plane of the shock (i.e., the component in one direction is much larger than the component in the orthogonal direction).
\end{itemize}

We divide the presentation of the results into two sections: isotropic and anisotropic pitch angle distribution.
We will perform calculations in the spectral range $\nu_m<\nu<\nu_c$, i.e., in the range of frequency between the injection frequency $\nu_m$ and the cooling frequency $\nu_c$. In this case the slope of the election distribution is equal to the slope of the injected distribution, since electrons are in slow cooling regime: $\delta=p$. We chose the value $p=2.5$, that implies a spectral index $s=(p-1)/2=0.75$ and $P^{lin}_{max}=0.72$.\\

%====================================  ISOTROPIC  ALPHA  ======================================
\subsection{Isotropic pitch angle distribution}\label{sect:isoa}
In case of isotropic pitch angle distribution, estimates of linear polarization have been already performed for all the magnetic field geometries that we are considering \citep{gruzinovwaxman99,ghisellini99,sari99b,granotkonigl03,granot03,nakar03,rossi04}. 

We report the results for completeness and for comparison with our results on circular polarization, since we are also interested in estimating the ratio $P^{circ}/P^{lin}$.

If the pitch angle distribution is isotropic then $dY/d\varphi=0$, and the equation for $P^{circ}_{0}$ simply reduces to:
%-------------------------------
\begin{equation}
P^{circ}_{0}= -f(\delta)P_{max}^{lin}\frac{\cot(\varphi)}{\gamma_e} 
\label{eq:pcirc_iso}
\end{equation}
%-------------------------------
\noindent
The function $f(\delta)$ depends very weakly on $\delta$, and assumes the value $f(\delta)\simeq2$ in the range $\delta=[2.1-2.8]$.
$\varphi$ and $\gamma_e$ as a function of the position on the map can be derived through equation \ref{eq:varphi_unif},\ref{eq:varphi_rad} and \ref{eq:ge}. However, the estimate of $\gamma_e$ depend on $\Gamma$ and $B$. In order to perform calculations which are independent on the specific values of these two parameters, instead of estimating $P^{circ}_{0}$ we estimate the quantity $P^{circ}_{0}\times\gamma_{e,min}$, where $\gamma_{e,min}$ is the minimum random Lorentz factor of those electrons contributing to the emission at frequency $\nu$. As it will become evident later, $P^{circ}_{0}\times\gamma_{e,min}$ does not depend on $\Gamma$ and $B$.
The relevant value of $\gamma_{e,min}$ can be estimated from the equations that we will give and by deriving the values of $\Gamma$ and $B$ from standard afterglow equations.
Note that, since the local circular polarization is of the order of $1/\gamma_e$, the total polarization will always be of the same order or smaller, no matter the geometry of the emission region (e.g., spherical or conical, top hat or structured jet, with uniform or angle dependent luminosity).

%===================================   ISO ALPHA -- UNIFORM B   =====================================
\subsubsection{A uniform magnetic field in the plane of the shock}\label{sect:isoa_unib}
%------------------------------------------------------
\begin{figure}
\hskip 0.6 truecm
\includegraphics[scale=0.66]{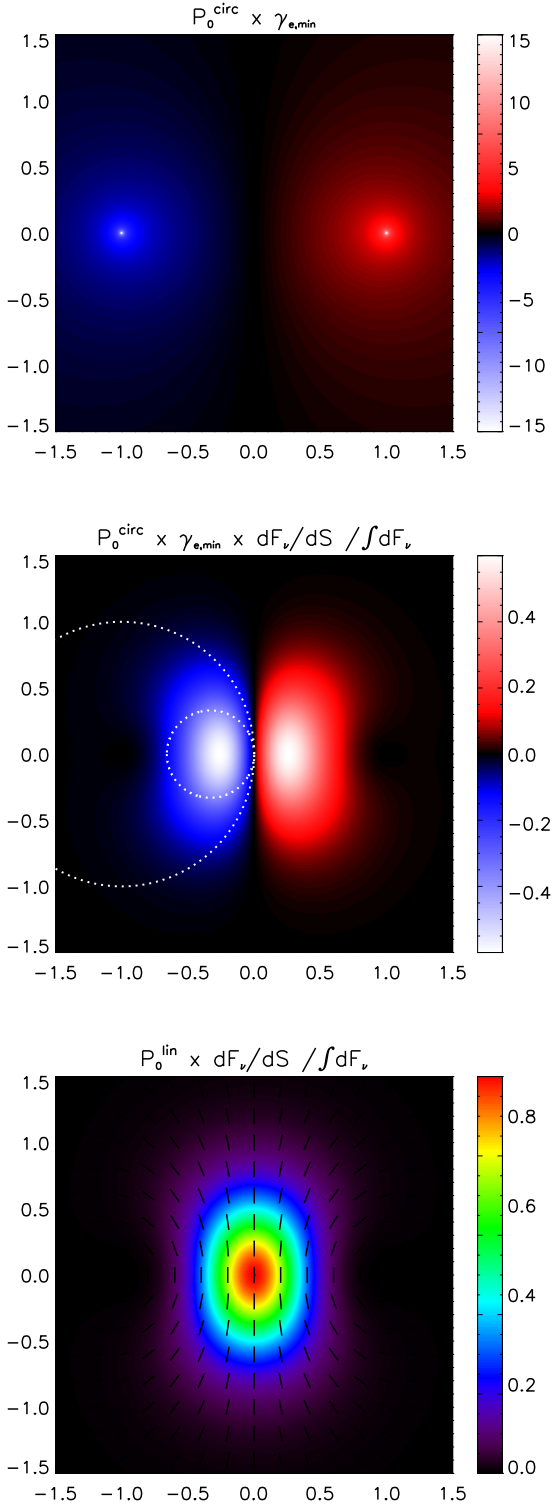}
\caption{Polarization levels for the case of an
isotropic pitch angle distribution and a uniform magnetic field in the plane of the shock.
The coordinate system $(x,y)=(r\cos\phi,r\sin\phi)$ adopted to build the maps is shown in Fig.~\ref{fig:sketch}.
Note that $r=1$ corresponds to an angular size $1/\Gamma$.
Upper panel: shown is the maximum local circular polarization $P^{circ}_0 \times \gamma_{e,min}$. Middle panel: flux-weighted maximum local circular polarization per unit surface. Bottom panel: flux weighted local linear polarization per unit surface. 
The dotted circles show the position of the jet needed to maximise the total circular polarization for a jet with $\theta_{jet}=1/\Gamma$ and $\theta_{jet}=1/3\Gamma$.}
\label{fig:isoa_unib}
\end{figure}
%------------------------------------------------------
We now consider a uniform magnetic field in the plane of the shock. The equations given below and the polarization maps are derived for a horizontally oriented (from left to right) magnetic field. 
In this configuration the angle $\varphi$ between the field and the photon (estimated in the comoving frame) is given by equation~\ref{eq:varphi_unif}.
The Lorentz factor $\gamma_e$ of electrons radiating at frequency $\nu$ reaches its minimum value when $y=0$, that is for electrons that move towards the observer (i.e., $\theta=0$). In this case $\varphi=90^\circ$ and $\sin\varphi=1$ (see equation~\ref{eq:ge}):

%-------------------------------------------
\begin{equation}
\gamma_{e,min}=\sqrt{\frac{\nu}{3\nu_H\Gamma}}=\gamma_e\sqrt{\frac{\sin\varphi}{1+y}}.
\end{equation}
%-------------------------------------------
From this last equation and eq.~\ref{eq:pcirc_iso} we obtain:
%-------------------------------------------
\begin{equation}
P^{circ}_0\times\gamma_{e,min}=-\sqrt{\frac{\sin\varphi}{1+y}}\cot{\varphi}\,f(\delta)P_{max}^{lin}.
\end{equation}
%-------------------------------------------

Figure~\ref{fig:isoa_unib} shows the polarization maps for the case of a uniform field.
The upper panel is the map of $P^{circ}_{0}\times\gamma_{e,min}$. Negative and positive values refer to the rotation direction of the polarization. For $\phi=\pm90^\circ$ the polarization is zero, since those photons that reach the observer are always perpendicular to the magnetic field ($\varphi=90^{\circ}$, see Figure~\ref{fig:sketch2}). If $\phi=0 \rm~(or~180^\circ)$ and $y=1$ then $\varphi=0,180^\circ$, and equation \ref{eq:p0circ} can not be applied. When $-90^\circ<\phi<90^\circ$, the angle between the magnetic field and the photon is $\varphi>90^\circ$ and the polarization is positive, while it is negative in the opposite case (see the discussion in Section \ref{sect:origin}). 

The central panel shows the map of the flux-weighted circular polarization $(P^{circ}_0\times\gamma_{e,min})_{fw}$ estimated from equation~\ref{eq:fwsp}, where $U_0/I_0$ in this case must be replaced by $P^{circ}_0\times\gamma_{e,min}$.

Because of left-right symmetry, in the case of a sphere, or a jet seen on-axis, or an off-axis jet with axis located at $\phi=\pm90^{\circ}$ the total integrated polarization $P^{circ}=0$, i.e., in this magnetic field configuration the total circular polarization vanishes regardless of what the local values of circular polarization are.
The integrated polarization can differ from zero only for an off-axis jet, and its value depends on $\theta_{jet}$ and on the position of the jet axis with respect to the observer.
We study the case of $\theta_{jet}\Gamma=1$ and $\theta_{jet}\Gamma=1/3$.
polarization maps are similar to those obtained for the spherical case, except for the fact that outside the region of the jet the flux is zero. 
We derived the total polarization for each different position of the jet axis in the plane of the sky and report in table~\ref{tab:iso_summary} its minimum and maximum value and the position at which these values are reached. 
The minimum polarization is zero, as previously discussed. The maximum polarization instead is reached when $\phi=0, 180^{\circ}$ and $r=0.3$ ($r=1$) for the narrower (larger) jet. The position of the jet needed to observe the maximum polarization is shown in Figure~\ref{fig:isoa_unib} for the two cases $\theta_{jet}\Gamma=1$ and $\theta_{jet}\Gamma=1/3$ with a dotted circle.

The bottom panel of Figure~\ref{fig:isoa_unib} shows the map of the flux-weighted linear polarization (before weighting for the flux the linear polarization is the same in each point of the map and is equal to $P^{lin}_{max}$). 
The total polarization in the spherical case is $P^{lin}=0.61$. 
Values of $P^{lin}$ at different frequencies, for different values of $p$ and for integration over $y_{max}=1$ and $y_{max}\gg1$ are reported in \cite{granot03}.

%====================================   ISO ALPHA -- RANDOM B  =================================
\subsubsection{A random magnetic field in the plane of the shock}\label{sect:isoa_randb}
The local circular polarization always vanishes for an isotropic pitch angle distribution. 
The linear polarization locally can be very high (see equation \ref{eq:plin0_random}), but due to the symmetry in this configuration, the total polarization vanishes for spherical blast waves and for wide jets. Net total linear polarization only arises for jets where the edges are visible.

%======================================   ISO ALPHA -- RADIAL B  =====================================
\subsubsection{A radial magnetic field}
\label{sect:isoa_radb}
In a radial magnetic field (i.e., perpendicular to the plane of the shock) $\mathbold B$ is parallel to the direction of the fluid velocity $\mathbold \beta$, and $\varphi=\theta^{\prime}$, leading to equation~\ref{eq:varphi_rad}.

%------------------------------------------------------
\begin{figure}
\hskip 0.6 truecm
\includegraphics[scale=0.66]{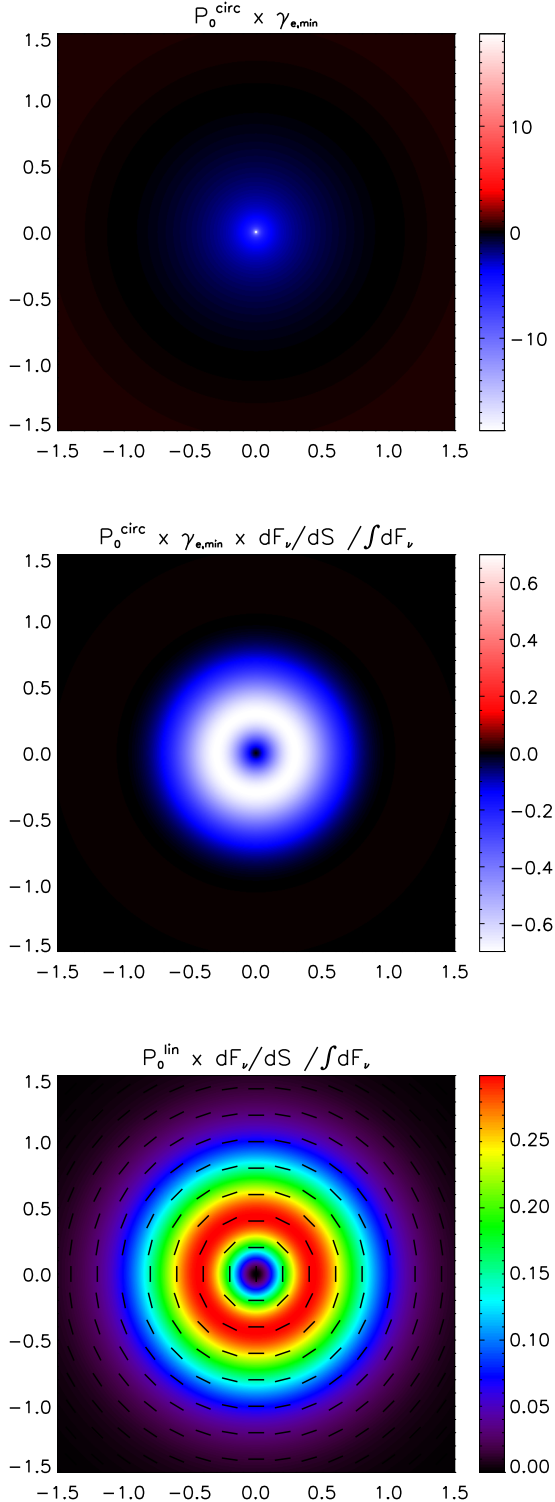}
\caption{
Polarization levels for the case of an
isotropic pitch angle distribution and a radial magnetic field perpendicular to the plane of the shock. 
The coordinate system $(x,y)=(r\cos\phi,r\sin\phi)$ adopted to build the maps is shown in Fig.~\ref{fig:sketch}.
Note that $r=1$ corresponds to an angular size $1/\Gamma$.
Upper panel: shown is the maximum local circular polarization $P^{circ}_0\times\gamma_{e,min}$. Middle panel: flux-weighted maximum local circular polarization per unit surface. Bottom panel: flux-weighted local linear polarization per unit surface.}
\label{fig:isoa_radb}
\end{figure}
%------------------------------------------------------
In this configuration the angle $\varphi$ does not depend on $\phi$.
The minimum electron Lorentz factor is reached when $y=1/3$.
The circular polarization (multiplied by $\gamma_{min}$) is:
%------------------------------------------------------
\begin{equation}
P_{0}^{circ}\times\gamma_{e,min}=-\frac{4(3y)^{1/4}}{3(1+y)}\cot{\varphi}\,f(\delta)P_{max}^{lin}.
\end{equation}
%------------------------------------------------------
The map of this geometry is shown in the upper panel of Figure~\ref{fig:isoa_radb}, while the middle panel shows the same quantity weighted by the flux. 
For $y<1$ (i.e., the region that mainly contributes to the observed flux) the angle $\varphi$ is always smaller than $90^{\circ}$ and the polarization is negative. 
Its contribution to the total polarization is not canceled by the positive contribution coming from the region at $y>1$, and the total polarization is $P^{circ}=-0.70/\gamma_{e,min}$. Jetted geometries can help to increase a bit this value (a factor of 2-3), but since locally the circular polarization is small, no large polarization can be reached by integration over the map (see table \ref{tab:iso_summary}).

Locally, the linear polarization is equal to $P^{lin}_{max}\simeq70\%$, but when integrated over all the emission region the total linear polarization is null: $P^{lin}=0$. This can be understood by looking at the orientation of the polarization vector in Figure~\ref{fig:isoa_radb} (bottom panel).
If jetted geometries are considered, then $P^{lin}$ can differ from zero and reach values as high as $40\%$ for very narrow jets.

From these considerations we conclude that for a radial magnetic field the ratio $P^{circ}/P^{lin}$ can go from $1/\gamma_e$ to infinitive. However, the absolute value of $P^{circ}$ is still limited to be $\lesssim 1/\gamma_e$.

%--------------------------------------
\begin{table*}
\begin{center}
\begin{tabular}{|c|c|c|c|c|c|c|}
\cline{3-7}
\multicolumn{2}{c|}{} & \multicolumn{2}{c|}{UNIFORM B}& {RANDOM B}& \multicolumn{2}{c|}{RADIAL B} \\
\cline{3-7}
\multicolumn{2}{c|}{} & $P^{lin}$ &$P^{circ}\times\gamma_{e,min}$ & $P^{lin}$ & $P^{lin}$ & $P^{circ}\times\gamma_{e,min}$ \\
\hline
\multicolumn{2}{|c|}{Sphere}  & 0.61 & 0 & 0 & 0 & 0.70\\
\cline{1-7}
\multirow{4}{*} {$\theta_j=1/\Gamma$} & \multirow{2}{*}{min} & 0.61  & 0 & 0 & 0 & 0.81\\
 & &($\phi=0,180^\circ; r=1$) & ($\phi=\pm 90^\circ$) & ($r=0$) & ($r=0$) & ($r=1$)\\
\cline{2-7}
 & \multirow{2}{*}{max} & 0.66 & 0.66  & 0.10  &0.20 & 1.0\\
 & & ($0.1<r<0.85$) & ($\phi=0,180^\circ; r=1$) && ($r=1$) & ($0.49<r<0.63$)\\
\cline{1-7}
 \multirow{4}{*} {$\theta_j=1/3\Gamma$} & \multirow{2}{*}{min} & 0.71  &0 & 0   &0 & 1.6\\
 & & & ($\phi=\pm90^\circ$) & ($r=0$) & ($r=0$) & ($r=0.33$)\\
\cline{2-7}
& \multirow{2}{*}{max} & 0.72 & 0.64& 0.10  &0.40  & 2.4 \\
& & & ($\phi=0,180^\circ; r=0.33$) &  & ($r=0.33$) & ($r=0$)\\
\hline
\end{tabular}
\end{center}
\caption{Isotropic pitch angle distribution: total linear and circular polarization for different geometries of the emitting region and different configurations of the magnetic field. The reported values for the circular polarization are multiplied by $\gamma_{e,min}$. For jetted geometers the polarization depends on the position ($r,\phi$) of the jet axis with respect to the observer, located at $r=0$. We list the maximum and minimum polarization that can be reached, and report the corresponding position of the jet. When $r$ and/or $\phi$ are not specified, it means that the value is independent on that parameter. In a random magnetic field circular polarization is always zero.}
\label{tab:iso_summary}
\end{table*}

%==============================================================================================
%====================================   ANISOTROPIC  ALPHA   =====================================
\subsection{Anisotropic pitch angle distribution}\label{sect:anisoa}
%-------------------------------------------------------------
\begin{figure}
\vskip -4.5 truecm
\hskip -6.2truecm
\includegraphics[scale=0.6]{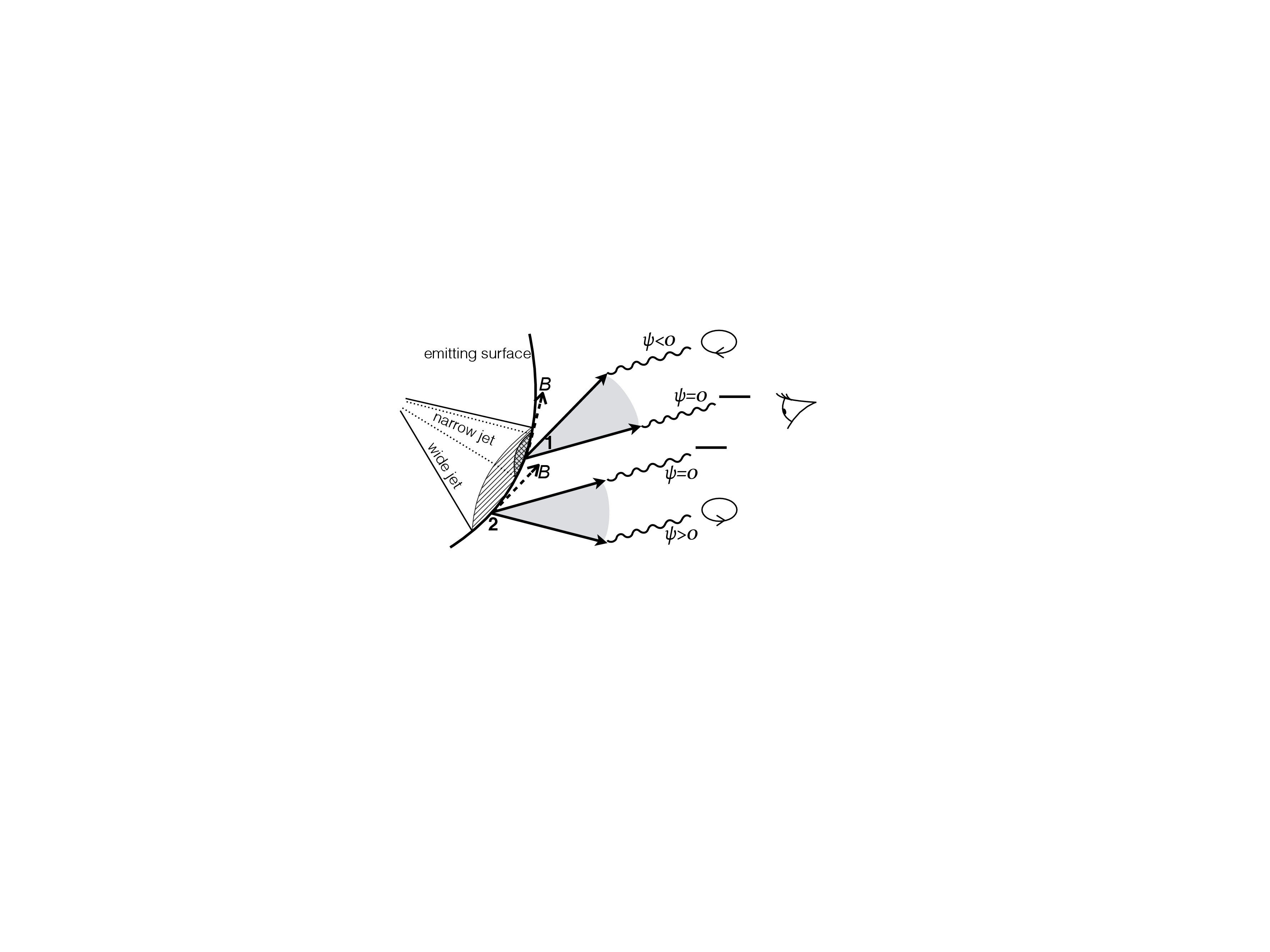}
\vskip -6.5 truecm
\caption{Schematic representation of the cancellation arising from different portions of the emitting region when an anisotropic pitch angle distribution is considered. In this example it is assumed that only electrons with pitch angles between $30^\circ$ and $60^\circ$ (shaded region) exist. The solid arrows show the edges of the distribution, while the dashed arrows show the magnetic field (assumed to be uniform in the plane of the shock). From region '1' the observer detects positive circular polarization, whose value can be high, due to the lack of electrons that would contribute with negative polarization. Conversely, from region '2' only electrons that contribute with a negative polarization are present. Outer regions do not contribute to the flux, since they are not emitting photons in the direction of the observer.}
\label{fig:sketch3}
\end{figure}
%--------------------------------------------------------------
For an isotropic pitch angle distribution, a large circular polarization (i.e., in excess of $1/\gamma_e$) cannot be achieved. This statement is true not only for the total integrated polarization, but also for a point-like region. In fact, as shown in Figure~\ref{fig:sketch2}, radiation originating from a point-like region is emitted by electrons with different pitch angles, and partial cancellation suppresses the polarization. However, if electrons are missing for some pitch angles (namely, if the angle distribution is anisotropic) then the polarization from a point-like region can be higher. For example (see Figure~\ref{fig:sketch2}), if all the electrons with pitch angle $\alpha<\varphi$ are missing, then the contribution from photons with $\psi>0$ is not suppressed
by contribution from photons with $\psi<0$. Then the condition for high $P^{circ}_0$ is that the observer is located just at the edge of the pitch angle distribution and that the level of anisotropy is very large. However, this is true if we consider the emission coming only from a specific location. Cancellation can still arise due to contribution from radiation emitted at different locations.

Consider two different point-like regions as in Figure~\ref{fig:sketch3} and imagine that the magnetic field is uniform in the plane of the shock (dashed arrows), and that electrons can only have pitch angles in the range $30^\circ-60^\circ$ (shaded areas). With reference to the point-like region '1', the observer is located at one of the edges of the distribution (solid arrows), so that the net polarization is high (due to the high derivative of the distribution) and is positive. However, when we consider region '2', the observer now sees the opposite edge of the distribution and the net polarization is still high, but negative. If we consider intermediate regions between '1' and '2' the polarization varies from large positive values down to zero, and then becomes negative with increasing absolute values, and reaches the maximum negative value at the location denominated '2'. Outside the region '1' to '2' the amount of radiation that reaches the observer is negligible. The only way to avoid cancellation from the positive and negative regions is to consider a jetted geometry. If the jet is large enough that all the region from '1' to '2' is within the jet surface (see 'wide jet' in Figure~\ref{fig:sketch3}), a strong cancellation takes place, like in the spherical case. However, if the jet is small and only one of the two regions is within the jet surface (see 'narrow jet' in Figure~\ref{fig:sketch3}) then the suppression is less significant and the integrated polarization can be larger.

From this qualitative discussion it is clear that an anisotropic pitch angle distribution can produce a high local polarization, but the polarization integrated over the emitting region will still be suppressed. If the visible region includes only the local regions characterised by high polarization, then the total polarization might be larger. However this requires a very narrow jet and/or fine tuned conditions on the location and extension of the jet. 
Another possibility to avoid such a cancellation is to consider either a emitting surface with no curvature or a patchy jet. In both cases, however, peculiar conditions between the pitch angle and the direction to the observer must be satisfied.
Summarising, from this qualitative discussion we conclude that high $P^{circ}$ can be achieved only in hydrodynamically unphysical configurations (such as a narrow jet or a planar geometry of the emission region) and/or in contrived and unrealistically fine tuned configurations.
Lorentz transformation (non considered in Figure~\ref{fig:sketch3}) do not changes the general results of this qualitative discussion.

In the following sections we present numerical estimates of the local and total polarization for different examples of anisotropic electron distributions and for different configurations of the magnetic field, and discuss the properties of the jet needed in order to maximise the circular polarization.

\subsubsection{Gaussian pitch angle distribution}
%------------------------------------------------------
\begin{figure}
\vskip -0.1truecm
\hskip 0.7 truecm
\includegraphics[scale=0.65]{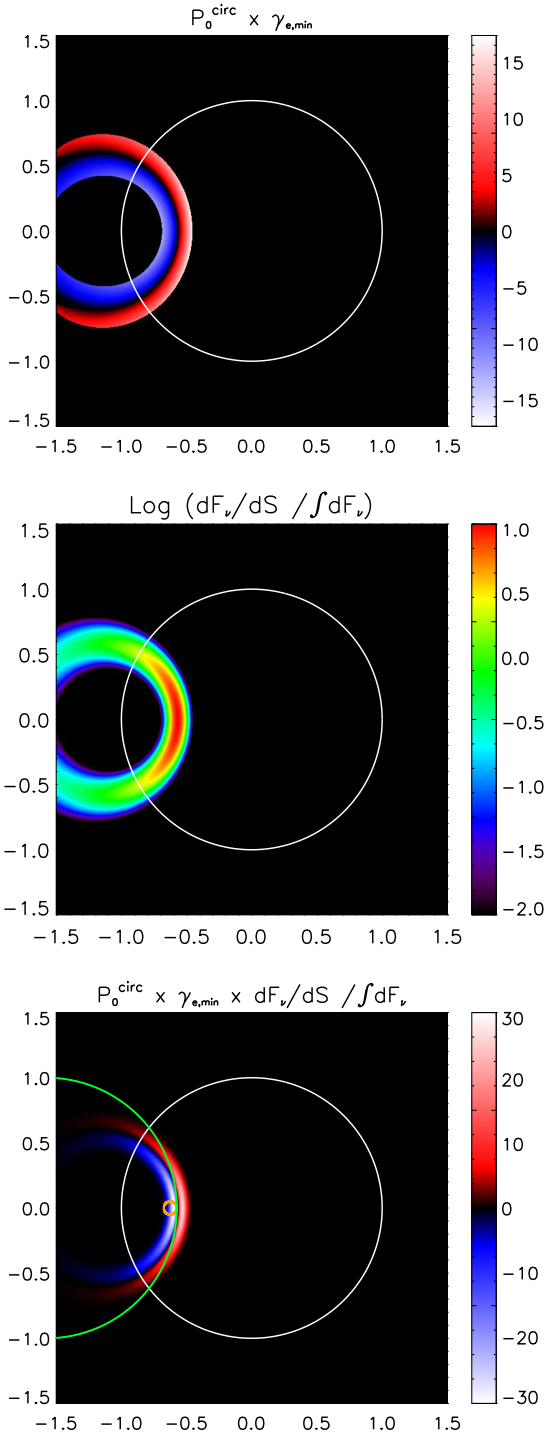}
\vskip -0.77truecm
\caption{
Polarization levels for
 the case of an anisotropic pitch angle distribution (a gaussian distribution peaked at  $\alpha_0=30^\circ$ and with width $\sigma=5\times10^{-2}$) and uniform magnetic field. 
Upper panel: maximum local circular polarization $P^{circ}_0\times\gamma_{e,min}$. Middle panel: weighted emissivity. Bottom panel: emissivity-weighted local circular polarization per unit surface. 
The positive and negative contributions nearly cancel each other and $P^{circ}\sim 0.3/\gamma_{e,min}$.
A higher polarization can be obtained by considering a jet whose position is selecting only (or mostly) the negative (or positive) region, like the ones represented by the green ($P^{circ}\sim 2/\gamma_{e,min}$) and orange ($P^{circ}\sim 5/\gamma_{e,min}$) circles.
The white solid circle shows the region located at $1/\Gamma$ from the observer.}
\label{fig:anisoa_unifb}
\end{figure}
%------------------------------------------------------

First, we consider a gaussian distribution $Y(\alpha)\propto \exp\left[-{\frac{(\alpha-\alpha_0)^2}{2\sigma_\alpha^2}}\right]$ and a uniform magnetic field in the plane of the shock.
The local polarization is derived from equation~\ref{eq:p0circ} and can reach very high values due to the factor $1/Y(\varphi)dY (\varphi)/d\varphi$. However, we need to account for the fact that the number of electrons with pitch angle $\alpha$ such that $|\alpha-\alpha_0|>\sigma_\alpha$ is negligible, and the flux from these regions is suppressed due to the lack of electrons with such pitch angles. Only electrons with $|\alpha-\alpha_0|\lesssim\sigma_\alpha$ contribute to the emission. Suppression of the emission at $|\alpha-\alpha_0|>\sigma_\alpha$ is assured by the multiplicative factor $Y(\varphi)$ in equation \ref{eq:wf}.
As a result, the radiation can reach the observer only from a small annular region for which the corresponding pitch angle is around $\alpha_0$. This can be clearly seen in the example in Figure~\ref{fig:anisoa_unifb} (where we have chosen $\alpha_0=30^\circ$ and $\sigma_\alpha=5\times10^2$).
As predicted by the qualitative discussion presented in the previous section, there is a region contributing with positive values and a region contributing with negative values.
In particular, for $\alpha > \alpha_0$ ($\alpha < \alpha_0$) the local polarization is positive (negative). As can be seen in Figure~\ref{fig:anisoa_unifb} (upper
panel) the maximum value is reached when the observer sees the edges of the distribution, i.e. for $|\alpha-\alpha_0|\simeq\sigma_\alpha$,
and its value is around $1/Y(\varphi)dY (\varphi)/d\varphi\sim1/\sigma_\alpha$. Thus the maximal circular polarization from a point like region is
$P^{circ}_{0,max}\sim1/(\sigma_\alpha\gamma_{e,min})$ (assuming $\sigma_\alpha\gamma_{e,min} > 1$, otherwise $P^{circ}_{0,max}\sim 1$). 
This explains why in the example in Figure~\ref{fig:anisoa_unifb} (where $\sigma_\alpha= 5 \times10^{-2}$), the maximum value of the local polarization (upper panel) is around  $\sim20/\gamma_{e,min}$. A larger polarization requires a narrower distribution, and implies a narrower extension of the two annular regions from where the radiation can reach the observer. 
The map for the flux weighted polarization is shown in the bottom panel and is the convolution of the map for the local circular polarization (upper panel) and the emissivity (middle panel).

When integration over the emitting surface is performed, partial cancellation from contributions with opposite rotation directions takes place. We found from numerical estimates that the final total polarization is of the order of $P^{circ} \lesssim1/\gamma_{e,min}$, and depends only slightly on the values of $\alpha_0$ and $\sigma_\alpha$ (for the particular case in Figure~\ref{fig:anisoa_unifb} we find $P^{circ} = 0.3/\gamma_{e,min}$). For $\alpha_0 = 90^\circ$ there is no net polarization, since the map is symmetric and perfect cancellation takes place. Strong cancellation can be avoided only by considering a jet 
whose position allows the observer to see only (or mostly) the region contributing with positive (or negative) polarization. 
This can be obtained, for example, with a very small jet with opening angle $\theta_{jet}\lesssim\sigma_\alpha/\Gamma$ (see Fig.~\ref{fig:anisoa_unifb}, bottom panel, orange circle).
Such a tiny jet is, however, hydrodynamically unphysical.
A wider jet may also result in a polarization much larger than that obtained in the spherical case, as long as the edge of the jet lies on the separation between the positive and negative region, as the one shown with a green circle in the lower panel of Fig.~\ref{fig:anisoa_unifb}.
This configuration requires, however, an extremely fine tuned position of the jet as compared to the observer.

Similar results can be obtained in the case of a radial magnetic field. In this case, the surface of equal pitch angles is an annular region centred around $r=0$, since $\varphi$ depends only on $r$ (see equation~\ref{eq:varphi_rad}), and goes from $\varphi=0^\circ$ to $\varphi=90^\circ$ going outward from $r=0$ to $r=1$. Again we have two annular concentric regions with opposite signs, one corresponding to the rising part of the pitch angle distribution, where the derivative is positive, and one corresponding to the decreasing part, where the derivative is negative. Contributions from these two regions cancel themselves and the total polarization is still of the order of $1/\gamma_e$. Again, we can avoid cancellation only if the jet contains one region but not the other one, which can be realised only if extremely fine tuned conditions are satisfied.

We conclude that a gaussian distribution can not produce a value of $P^{circ}$ in excess of $1/\gamma_e$ under realistic conditions.

\subsubsection{General distribution}
\begin{figure}
\vskip -4 truecm
\hskip -4 truecm
\includegraphics[scale=0.43]{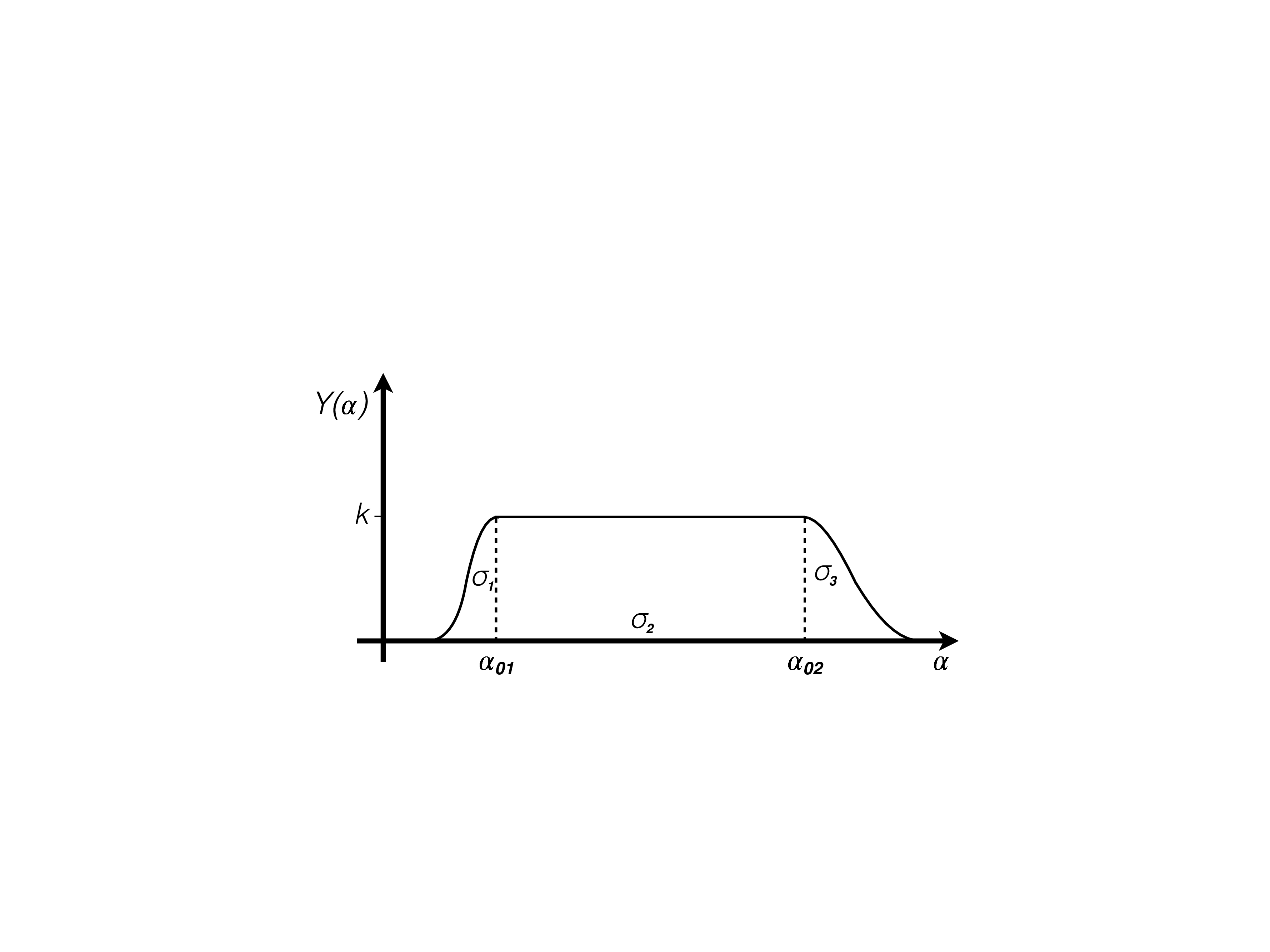}
\vskip -2.8 truecm
\hskip 0.2 truecm
\includegraphics[scale=0.74]{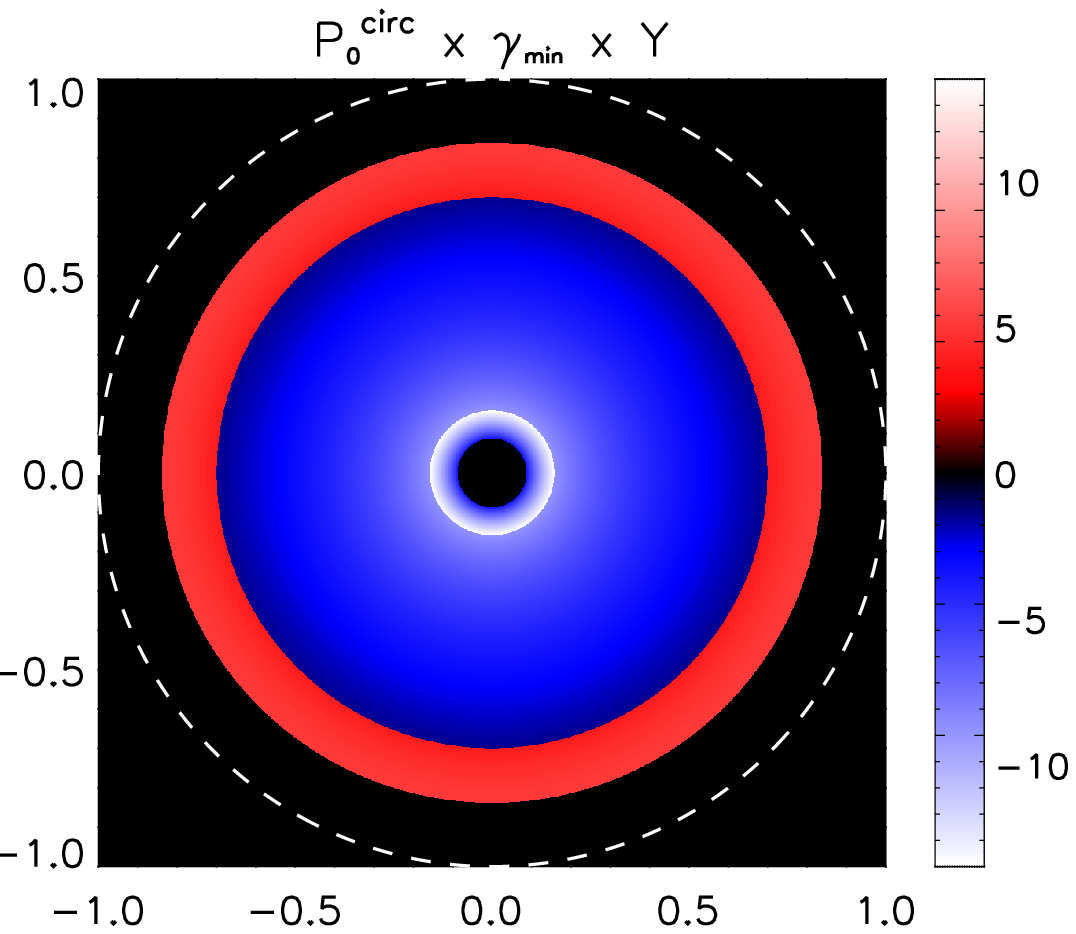}
\caption{The lower panel shows the circular polarization levels for the pitch angle distribution $Y(\alpha)$ depicted in the upper panel. The coordinate system $(x,y)=(r\cos\phi,r\sin\phi)$ adopted to build the maps is shown in Fig.~\ref{fig:sketch}.
 Region $\sigma_1$ corresponds to the inner annular region characterised by large and negative local polarization. Region $\sigma_3$ corresponds to the outer annular region with large and positive polarization. The two are separated by region $\sigma_2$. The dashed circle corresponds to $r=1$.}
\label{fig:general}
\end{figure}
%-----------------------------------------
While in the previous section we have presented numerical results for the specific case of a gaussian distribution, in this section we consider a more general distribution, and present order of magnitude estimates of the total circular polarization. We show that the numerical results derived in the previous section for the specific case of a gaussian distribution also apply to more general functions $Y(\alpha)$.

We consider a pitch angle distribution $Y(\alpha)$ (shown in figure~\ref{fig:general}, top panel) where the rising and decreasing part can have different widths ($\sigma_1$ and $\sigma_3$, respectively) and are separated by a region of width $\sigma_2$ where the distribution is flat (i.e., the number of electrons per solid angle is constant). An example of the polarization map corresponding to this distribution is shown in the bottom panel of figure~\ref{fig:general}, for the case of a radial magnetic field.

Neglecting numerical factors order unity, the local circular polarization is such that $P^{circ}\gamma_e\simeq\mp1/\sigma_{1/3}$ in the
rising and decreasing part of the distribution (region 1 and 3 respectively), and $P^{circ}\gamma_e\simeq-1$ in the flat part (region 2).
We assume that both $\sigma_1$ and $\sigma_3$ are much smaller than 1, so that in these two regions the local polarization is large. 
The total contribution coming from each region is then given by the local polarization multiplied by the area: $\Delta S_1/\sigma_1$, $\Delta S_2$, and $\Delta S_3/\sigma_3$, respectively for the three different regions. The gaussian distribution discussed in the previous section can be recovered for $\sigma_2=0$ and $\sigma_1=\sigma_3$. In this case, $P^{circ}\gamma_e \sim 1/\sigma\times(\Delta S_3-\Delta S_1)/(\Delta S_3+\Delta S_1)$. The suppression coming from integration over the two regions is of the order of $\sigma$, so that the total polarization is $P^{circ}\sim 1/\gamma_e$. In fact, the contribution from the two regions is of the same order of magnitude (but with opposite signs) and almost cancel each other. More specifically, the sum of the two areas is of the order of $r_0^2\sigma$, but their difference is of the order of $r_0^2\sigma^2$. The final result $P^{circ}\sim1/\gamma_e$ is valid both for $\sigma_1\approx\sigma_3$ and for $\sigma_1\ll\sigma_3$ (or viceversa). In the last case, indeed, the local polarization in region 1 is much higher than the local polarization in region 3, but the area is much smaller, by the same factor. 
Again, the two contributions are of the same order of magnitude and cancel each other.

One can wonder if it is possible to avoid cancellation between the two terms by having $\Delta S_1<<\Delta S_3,$ but $\sigma_1\gtrsim\sigma_3$, so that the contributions from the two regions are now very different. This configuration can be achieved by locating the two regions far from each other. This means to consider an intermediate region where the distribution is flat over a wide range of pitch angles: $\sigma_2\gg\sigma_{1/3}$. In this case $\Delta S_1/\sigma_1\ll \Delta S_3/\sigma_3$ and can be neglected. The two regions characterised by large opposite polarization have now very different areas and do not cancel each other (see for example figure~\ref{fig:general}). However, the total flux is dominated by region 2, and $P^{circ}\sim1/\gamma_e(\Delta S_3/\sigma_3 -\Delta S_2)/\Delta S_2\sim 1/\gamma_e$.

We have demonstrated that the total integrated polarization is always of the order of $1/\gamma_e$, no matter the presence
of regions characterised by a large local polarization. As already discussed in the previous section, a possibility to avoid
suppression is to have a jetted outflow whose extension and location allows to observer to see only photons with (large)
positive or negative polarization, i.e. to see only region 1 or 3. 
Let's call $\sigma\ll1$ the width of this region. If $\theta_{jet}\Gamma\lesssim\sigma$ then $P^{circ}\lesssim\frac{1}{\gamma_e\sigma}$. We stress that this result is valid only if $\theta_{jet}\Gamma\lesssim\sigma\ll1$, and for a fine tuned location of the jet with respect to the observer.

\section{Discussion: the circular polarization in GRB~121024A}
\label{sect:discussion}
Polarization measurements in the afterglow of GRB~121024A revealed that 0.15 days after the burst the optical radiation was circularly polarised at a level of 0.6\%, and linearly polarised at a level of 4\%, implying a ratio $P^{circ}/P^{lin}\simeq0.15$ \citep{wiersema14}. The claim of the detection was accompanied by the claim that these numbers are much in excess of the theoretically expected values of $1/\gamma_e$, since the estimated $1/\gamma_e$ for electrons radiating in the optical at the time of the detection is $1/\gamma_e\sim10^4$. Anisotropies in the pitch angle distribution have then been invoked as a possible explanation \citep{wiersema14}.

Our studies imply that, no matter the pitch angle distribution, if optically thin synchrotron radiation is dominating the emission, values of $P^{circ}$ much in excess of $1/\gamma_e$ can be achieved only 
in very contrived and/or unphysical geometries, that require extremely fine tuned conditions.
We are then left with the possibility that $\gamma_e$ is small, so that $P^{circ}\sim1/\gamma_e$ can be large enough to explain the observations. We then revisit the estimate of $\gamma_e$ to understand if this is a viable possibility.

The minimum Lorentz factor of those electrons that mainly contribute to the observed optical frequency $\nu_{obs} = 4.6\times10^{14}\,$Hz is $\gamma_{e,min} =\sqrt{2\pi m_e c(1+z)\nu^{obs}/(3\Gamma Bq_e)}$. Using the self-similar Blandford \& McKee solution to estimate $\Gamma$ and using $B =\sqrt( 32\pi\epsilon_Bnm_pc^2\Gamma^2)$ (with $n = n_0 = const$ for a homogeneous medium and $n = 3\times10^{35}A_\star R^{-2}$ for a wind-like medium) we find respectively:
%---------------------------
\begin{eqnarray}
\gamma_{e,min}&=&3\times10^3 E_{K,52}^{-1/8}n_0^{-1/8}\epsilon_B^{-1/4} \qquad\rm (ISM),\\
\gamma_{e,min}&=&1.3\times \epsilon_B^{-1/4} A_\star^{-1/4} \qquad\qquad~~\rm (Wind).
\end{eqnarray}
%---------------------------
In order to explain the observed circular polarization, according to our numerical results, a value $\gamma_{e,min}\lesssim100$ is needed, implying:
%---------------------------
\begin{eqnarray}
E_{K,52}^{1/8}n_0^{1/8}\epsilon_B^{1/4}>30 \qquad\rm (ISM),\\
\epsilon_B^{1/4} A_\star^{1/4}>13 \qquad\rm (Wind).
\end{eqnarray}
%---------------------------
Even for a large $\epsilon_B\sim1$, the first condition can hardly be
satisfied and the ISM case is ruled out, while in the wind-like case it is possible to account for a small $\gamma_{e,min}\sim100$ provided that the density is large: $A_\star > 3\times10^4$. Even admitting the possibility to have such a strong wind from the precursor star, this large density would imply a transition to non-relativistic velocity at early times, when the medium is still optically thick and no radiation can be emitted.

In the context of optically thin synchrotron radiation, the possibility to explain $P^{circ}\sim0.6\%$ in GRB 121024A as originated by electrons with small Lorentz factor is then ruled out.

\section{Conclusions}
\label{sect:conclusions}
In this work we have estimated the level of circular polarization of optically thin synchrotron radiation from a spher- ically expanding extended source moving at relativistic velocity. The equations for the estimate of circular polarization in synchrotron radiation from a point-like region in the frame of the fluid have been derived by \cite{legg68}, \cite{sazonov69}, and \cite{sazonov72}. However, in order to apply these equations to the study of emission from relativistic sources it is necessary i) to account for transformations from the fluid frame to the observer frame, where the emitting fluid is moving at relativistic velocity, and ii) to perform integration over the whole visible emitting region. We have considered these two effects and presented the equations that should be applied to derive $P^{circ}$ from a relativistic unresolved source.

We have considered three different configurations of the magnetic field: i) uniform in the plane of the shock, ii) radial, iii) random in the plane of the shock. In this latter case, $P^{circ}$ vanishes in each point of the emitting region. In a uniform magnetic field, even if the local polarization is of the order of $1/\gamma_e$, integration over the emitting region plays a fundamental role: for spherical outflow (or for wide jet or smaller jets seen on-axis) the symmetry of the system suppresses $P^{circ}$, that vanishes due to the cancellation between regions with different directions of the polarization rotation. Values different from zero can be obtained only when the observer sees the edges of the jet and the jet is off-axis. In this case, values up to $1/\gamma_e$ can be recovered. In a radial magnetic field, instead, the region inside the cone $1/\Gamma$ contributes with negative polarization, while the region outside (which however gives a negligible contribution to the observed flux) contributes with positive polarization. In this case, the total value differs from zero also for the spherical/wide jet case. However, also in this case its value does not exceed the limit $1/\gamma_e$. To summarize, for isotropic pitch angle distributions the value of $P^{circ}$ is limited to be smaller than $1/\gamma_e$, and Lorentz transformations and integration over the surface might reduce the observed level of polarization, that is always similar or smaller than the one estimated for a point-like source observed in its comoving frame.

For all the considered configurations we have also de- rived the value of the ratio $P^{circ}/P^{lin}$. Contrary to what is generally stated, the ratio between circular and linear polarization is not necessarily of the order of $1/\gamma_e$, but can indeed assume any value, depending on the configuration of the magnetic field and on the geometry of the emitting region.

We have investigated the possibility to observe a high level of $P^{circ}>1/\gamma_e$ as a result of an anisotropic distribution of the electron pitch angles, motivated by the recent detection in the optical afterglow of GRB 121024A. We have demonstrated that invoking a large anisotropy does not help to overcome the limit $1/\gamma_e$. The only difference between the isotropic and anisotropic case is the following. For an isotropic distribution, the circular polarization of radiation coming from a point-like region is strongly suppressed, due to average over electrons with different pitch angles. For an anisotropic distribution the lack of electrons with some pitch angles can limit the cancellation and give rise to high values of local polarization. However, cancellation now arises from the integration over the emitting region. The numerical results and approximate analytic calculations presented in this work show that after integration, the total circular polarization is still of the order of $1/\gamma_e$. A higher $P^{circ}$ 
can be reached only by  
considering a jet satisfying unphysical conditions (as a planar emitting region or a tiny jet) and/or very unlikely conditions, as an extremely fine tuned location of the jet as compared to the observer.
The application of our study to GRB 121024A leads us to conclude that the circular polarization measured
in the optical afterglow of this GRB cannot be explained in the context of optically thin synchrotron radiation, no matter the configuration of the magnetic field and the shape of the pitch angle distribution.

\section*{Acknowledgements}
This research was supported by the I-CORE Program of the PBC and the ISF (grant 1829/12). LN was supported by a Marie Curie Intra-European Fellowship of the European Community's 7th Framework Programme (PIEF-GA-2013-627715). TP was partially supported by a China-Israel Collaboration and by an ISA grant. EN was partially supported by an ERC starting grant (GRB/SN), ISF grant (1277/13) and an ISA grant.

\bibliography{biblio.bib}

\begin{thebibliography}{}

\bibitem[\protect\citeauthoryear{{Covino}, {Ghisellini}, {Lazzati} \&
  {Malesani}}{{Covino} et~al.}{2004}]{covino04}
{Covino} S.,  {Ghisellini} G.,  {Lazzati} D.,    {Malesani} D.,  2004, in
  {Feroci} M.,  {Frontera} F.,  {Masetti} N.,   {Piro} L.,  eds, Gamma-Ray
  Bursts in the Afterglow Era Vol.~312 of Astronomical Society of the Pacific
  Conference Series, {Polarization of Gamma-Ray Burst Optical and Near-Infrared
  Afterglows}.
p.~169

\bibitem[\protect\citeauthoryear{{Ghisellini} \& {Lazzati}}{{Ghisellini} \&
  {Lazzati}}{1999}]{ghisellini99}
{Ghisellini} G.,  {Lazzati} D.,  1999, \mnras, 309, L7

\bibitem[\protect\citeauthoryear{{Granot}}{{Granot}}{2003}]{granot03}
{Granot} J.,  2003, \apjl, 596, L17

\bibitem[\protect\citeauthoryear{{Granot} \& {K{\"o}nigl}}{{Granot} \&
  {K{\"o}nigl}}{2003}]{granotkonigl03}
{Granot} J.,  {K{\"o}nigl} A.,  2003, \apjl, 594, L83

\bibitem[\protect\citeauthoryear{{Gruzinov}}{{Gruzinov}}{1999}]{gruzinov99}
{Gruzinov} A.,  1999, \apjl, 525, L29

\bibitem[\protect\citeauthoryear{{Gruzinov} \& {Waxman}}{{Gruzinov} \&
  {Waxman}}{1999}]{gruzinovwaxman99}
{Gruzinov} A.,  {Waxman} E.,  1999, \apj, 511, 852

\bibitem[\protect\citeauthoryear{{Legg} \& {Westfold}}{{Legg} \&
  {Westfold}}{1968}]{legg68}
{Legg} M.~P.~C.,  {Westfold} K.~C.,  1968, \apj, 154, 499

\bibitem[\protect\citeauthoryear{{Mundell}, {Kopa{\v c}}, {Arnold}, {Steele},
  {Gomboc}, {Kobayashi}, {Harrison}, {Smith}, {Guidorzi}, {Virgili}, {Melandri}
  \& {Japelj}}{{Mundell} et~al.}{2013}]{mundell13}
{Mundell} C.~G.,  {Kopa{\v c}} D.,  {Arnold} D.~M.,  {Steele} I.~A.,  {Gomboc}
  A.,  {Kobayashi} S.,  {Harrison} R.~M.,  {Smith} R.~J.,  {Guidorzi} C.,
  {Virgili} F.~J.,  {Melandri} A.,    {Japelj} J.,  2013, \nat, 504, 119

\bibitem[\protect\citeauthoryear{{Nakar}, {Piran} \& {Waxman}}{{Nakar}
  et~al.}{2003}]{nakar03}
{Nakar} E.,  {Piran} T.,    {Waxman} E.,  2003, \jcap, 10, 5

\bibitem[\protect\citeauthoryear{{Rossi}, {Lazzati}, {Salmonson} \&
  {Ghisellini}}{{Rossi} et~al.}{2004}]{rossi04}
{Rossi} E.~M.,  {Lazzati} D.,  {Salmonson} J.~D.,    {Ghisellini} G.,  2004,
  \mnras, 354, 86

\bibitem[\protect\citeauthoryear{{Rybicki} \& {Lightman}}{{Rybicki} \&
  {Lightman}}{1986}]{rybicki86}
{Rybicki} G.~B.,  {Lightman} A.~P.,  1986, {Radiative Processes in
  Astrophysics}

\bibitem[\protect\citeauthoryear{{Sari}}{{Sari}}{1999}]{sari99b}
{Sari} R.,  1999, \apjl, 524, L43

\bibitem[\protect\citeauthoryear{{Sazonov}}{{Sazonov}}{1969}]{sazonov69}
{Sazonov} V.~N.,  1969, \sovast, 13, 396

\bibitem[\protect\citeauthoryear{{Sazonov}}{{Sazonov}}{1972}]{sazonov72}
{Sazonov} V.~N.,  1972, \apss, 19, 25

\bibitem[\protect\citeauthoryear{{Westfold}}{{Westfold}}{1959}]{westfold59}
{Westfold} K.~C.,  1959, \apj, 130, 241

\bibitem[\protect\citeauthoryear{{Wiersema}, {Covino}, {Toma}, {van der Horst},
  {Varela}, {Min}, {Greiner}, {Starling} \& {et al.}}{{Wiersema}
  et~al.}{2014}]{wiersema14}
{Wiersema} K.,  {Covino} S.,  {Toma} K.,  {van der Horst} A.~J.,  {Varela} K.,
  {Min} M.,  {Greiner} J.,  {Starling} R.~L.~C.,    {et al.} 2014, \nat, 509,
  201

\end{thebibliography}

\label{lastpage}

\end{document}